# Contents

# Cloud Fog Architectures in 6G Networks

Barzan A. Yosuf, Amal A. Alahmadi, T. E. H. El-Gorashi and Jaafar M. H. Elmirghani

**Abstract** prior to the advent of the cloud, storage and processing services were accommodated by specialized hardware, however, this approach introduced a number of challenges in terms of scalability, energy efficiency, and cost. Then came the concept of cloud computing, where to some extent, the issue of massive storage and computation was dealt with by centralized data centers that are accessed via the core network. The cloud has remained with us thus far, however, this has introduced further challenges among which, latency and energy efficiency are of the pinnacle. With the increase in embedded devices' intelligence came the concept of the Fog. The availability of massive numbers of storage and computational devices at the edge of the network, where some are owned and deployed by the end-users themselves but most by service operators. This means that cloud services are pushed further out from the core towards the edge of the network, hence reduced latency is achieved. Fog nodes are massively distributed in the network, some benefit from wired connections, and others are connected via wireless links. The question of where to allocate services remains an important task and requires extensive attention. This chapter introduces and evaluates cloud fog architectures in 6G networks paying special attention to latency, energy efficiency, scalability, and the trade-offs between distributed and centralized processing resources.

## Introduction

During the past several decades, computing paradigms have evolved from distributed models that included dedicated hardware such workstations to a more centralized model that is widely referred to as cloud computing. Cloud computing data centers are typically accessed via the Internet as they are attached to the core network [1]. This remote processing using a centralized cloud approach would not have been possible had it not been for the great advancement in both wired and wireless communication networks in terms of the speed at which data could be transmitted [2]. To some extent, cloud computing resolved the issue of massive storage and computation requirements of many applications. Cloud computing has two important traits [3]. First, centralization facilitates economies of scale through minimizing the cost of administration and operations. Second, speeding up innovations as individuals and organizations can avoid the operational and capital expenditure associated with owning a data center [3]. However, despite the cloud's on-demand and scalability merits, accessing its resources requires traversing through the access, metro and core network layers that can be prohibitively costly. This cost can be in terms of the communication latency due to the distance between data source nodes and the cloud and the high power consumption induced due to the transport network [4].

As a result, a new model of computing was proposed by Cisco in 2012, which is widely known as fog computing [2]. The term "fog" is used metaphorically to differ from the "cloud" as it is near to the ground [5]. In the same way, the main goal of fog computing is to extend the cloud services from the core to the edge of the network [6]. Thus, fog resources (i.e. computing and storage) are geographically distributed in the network through an *N-tier* deployment whereby heterogeneous fog resources are provided at different hierarchical levels [1]. According to the OpenFog consortium, any device can act as a fog node, be it embedded type CPUs onboard smart IoT devices or servers co-located with ISP's regional offices to processing servers located at local offices and/ or customer premises [7].

Driven by the emergence of the Internet of Things (IoT), the number of connected devices is expected to grow exponentially. Estimates have reported these devices to range between 25 and 50 billion, generating around 79.4 zettabytes of data [8], [9]. Evidently, given the rate at which connected devices grow, engineers from both industry and academia have already begun research into 6G networks, while 5G is currently being rolled out commercially. 6G networks are expected to support a new breed of next-generation applications (e.g. augmented reality, remote surgery, etc.) and an abundance of connectivity for the massive number of connected devices also referred to as supper IoT [10], [11]. One of the major aspects of 6G networks is integrating machine learning (ML) tools such as artificial intelligence (AI) for data analytics in order to move away from manual configurations / optimizations to a more intelligent network in the future [10].

As stated by one of the first white papers on 6G, a latency threshold of 0.1ms and an improvement of 10x in energy efficiency are among the most important key performance indicators (KPI) in 6G [12]. Therefore, investigating cloud fog architectures in future 6G networks is imperative, because if the massive amounts of data generated at the edge network were to be processed centrally by the cloud would lead to slow decision making due to latency and increased power consumptions due to the transport network [13]. Hence, the interplay between the fog and the cloud in terms of energy efficiency and latency will become an important aspect of 6G networks due to the increased use of processing and data analytics [14]. The centralized cloud data center (DC) can provide increased processing capabilities and sophistication but may result in higher power consumption and increased latency. Thus, resource allocation problems will become vital where specialized compute and storage hardware must be accessed either in the distant cloud or in the edge processing fog nodes. It is anticipated that through the complementary features and the cooperation between the fog and the cloud, a more efficient and greener network can be achieved [15].

Different from the works that address the resource provisioning problem in cloud fog architectures only on individual layers, this chapter introduces a comprehensive optimization model based on Mixed Integer Linear Programming (MILP) and extends on the work in [16] by paying special attention to latency as well as energy consumption. The optimization model in this chapter considers i) elements in the IoT, access, metro, and core layers, ii) simultaneous task requests generated from multiple IoT groups to capture the trade-offs between local and remote and/ or centralized processing, iii) a generic MILP model that is independent of the type of processing and/or networking technologies which allows for a holistic focus on energy-efficiency with a global perspective. Also, the work in this chapter benefits from our previous contributions in improving energy efficiency of cloud DCs [17] – [21], big data analytics [22] – [25], network coding in core networks [26], [27], energy efficient optical core networks [13], [28] – [34], and virtualization in the core and IoT networks [35] – [37].

# The Proposed Cloud Fog Architecture

Devices at the edge of the network such as user equipment, vehicles and smart IoT devices are expected to possess low power embedded and specialized CPUs, which can collectively provide enormous amounts of computational power due to their massive numbers and very low latencies due their distributed nature and proximity to end-users. The IoT and fog nodes are highly heterogeneous in terms of their processing resources and efficiencies, which poses a number of challenges in the resource allocation problem in future 6G networks. Therefore, the proposed cloud fog architecture shown in **Figure 1** comprises of 5 distinct layers of processing that comprise of the IoT, CPE, Access Fog, Metro Fog, and Cloud DC layers. A generated task emanates from the IoT layer and may be hosted by any of the aforementioned processing locations given that they have enough processing resources.

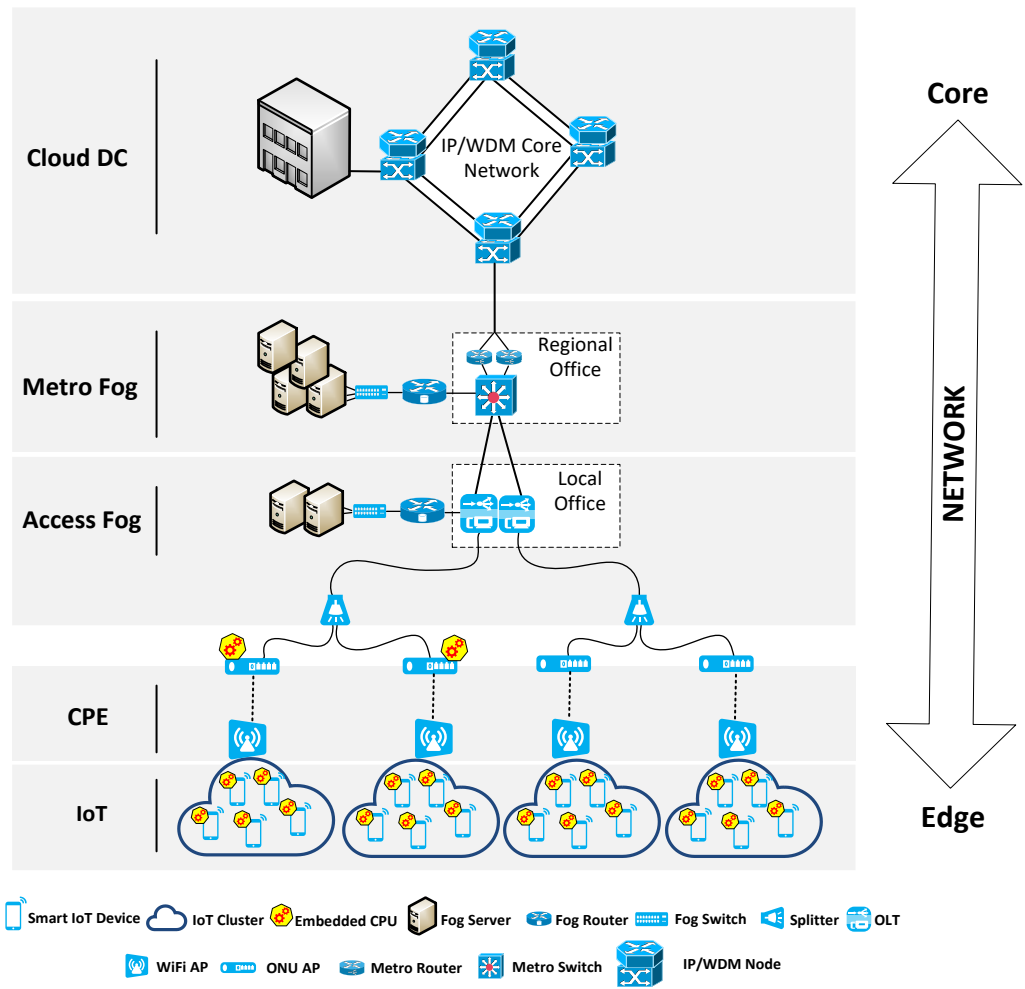

*Figure 1* A multi-layer cloud fog architecture supported by a PON access network.

A. IoT Layer

This is the bottom-most layer and comprises of all the generic smart IoT nodes such as tablets, phones, vehicles etc. Two types of entities are defined in this layer, which are called source nodes and IoT nodes. The sensor nodes are a subset of the IoT nodes, the only difference between them is that the former is generating task requests while the latter remains idle. A real-world example can include a smart surveillance system whereby one or more cameras actively send video streams while the rest of the cameras remain idle due to little or no motion detected by their mounted passive infrared (PIR) sensors. The IoT nodes are wirelessly connected to the wireless access points (APs) in their respective zones and the generated tasks from the source nodes are offloaded to the next layer(s) for data analysis, if local resources are insufficient. A Wi-Fi link is considered between the IoT and APs as it is an ideal choice for data-intensive applications compared with other wireless counterparts such as Bluetooth, Zigbee, LoRa, etc. In indoor environments, this link may also be replaced with a visible indoor light link in order to support very high data rates per user terminal [38].

B. CPE Layer

This layer comprises of the customer premises equipment (CPE) such as ONUs and Wi-Fi Access Points (APs). These devices are typically situated in proximity to the IoT nodes. The ONU, since it has multiple Ethernet ports and acts as a switch can be equipped with embedded type CPUs that have similar and /or higher capacity than the IoT nodes [39]. Small organizations or even end-users can deploy their own CPE nodes at locations such as APs, routers, gateways, etc. In this chapter, Optical Networking Units (ONUs) represent the CPE nodes and they are part of the Passive Optical Network (PON) deployment [40]. PON technologies have promising potentials as they offer high data rates for data-intensive applications, at relatively low cost and PONs are particularly suitable due to their high scalability [41]. The main role of the CPE nodes is to act as controllers that collect and coordinate the allocation of the generated tasks by the source nodes. Thus, these CPE nodes are assumed to be fully aware of the available processing resource across all layer of the architecture shown in **Figure 1**. As can be seen, each CPE node is connected to a separate IoT group which represents a geographical area. The CPE nodes due to their coordination and allocation roles can communicate with other group(s) through the PON access network. Thus, the tasks generated from one group can also be allocated to other zones [42]. The access part of the PON deployment will be explained next.

C. Access Fog Layer

This layer comprises of multiple Optical Line Terminals (OLTs) that are responsible for aggregating data traffic from the connected ONU devices. A single fiber link can be split in the ratio of 1:N and with next generation PONS (NG-PONs) a splitting ratio of 1:256 can be achieved [43]. This is particularly suitable for 6G networks as it is expected that there will be hundred(s) of devices per cubic meter [12]. The processing capability available to this layer is higher than that of the IoT and the CPE as several high-end servers are used to form a fog collocated with the OLT [44]. However, the number of servers that can be deployed at this layer is still finite, which can be due to space limitations as OLT devices are usually installed in small local offices and/or enclosed in street cabinets. Thus, more intensive tasks will need to be offloaded to the next layer for processing.

D. Metro Fog Layer

This layer comprises of multiple edge routers and a single ethernet switch that acts as the entry point to the metro and edge network as can be seen in **Figure 1**. The Ethernet switch is mainly used to provide access to public clouds and is also used for traffic aggregation from one or more access networks (OLT devices in this chapter). The main role of the edge routers is to perform traffic management and authentication and usually multiple edge routers are used for redundancy purposes [41]. The computational resources available to the metro fog layer are typically significantly higher than that of the IoT and lower fog layers due to the number of users it supports, however the resources are still insignificant compared to the cloud DC [45].

E. Cloud DC Layer

This layer comprises of large data centers that are attached to the core network. High capacity IP/WDM fibre links are used to interconnect the core nodes. The IP/WDM core network consists of an optical layer and an IP layer. In the IP layer, IP core routers are deployed at each node to aggregate traffic and / or route traffic. In the optical layer, optical cross connects are used to establish the physical network links between the IP core routers. WDM fiber links utilize EDFAs, transponders and regenerators as part of the IP/WDM setup. The processing resources of the cloud DC are virtually infinite in comparison to the rest of the aforementioned processing layers. This is understandable since cloud DCs are not restricted by space, they are deployed to support vast number of applications and service [46].

## MILP Model

The proposed cloud fog architecture is shown in **Figure 1**. The optimization model minimizes the joint networking and processing power consumption resulting from IoT processing placement [42], [70]. Each task request comprises of CPU and traffic requirement. CPU is the amount of processing required in Million Instructions Per Second (MIPS) and traffic is the amount of data required to be transported in the network in Mbps. The optimization model considers the network topology in **Figure 1** as a bi-directional graph $G(N, L)$, where $N$ is the set of all nodes, and $L$ is the set of links connecting those nodes. A processing node's computational capacity is measured in MIPS, whilst the link's network capacity is given in Mbps.

The definitions of the sets, parameters, and variables used in the MILP model are as follows:

**Sets:**

| | |
|---|---|
| $\mathbb{N}$ | Set of all nodes in the proposed architecture shown in Figure 1. |
| $\mathbb{N}_m$ | Set of all neighbor nodes of node $m$ in the proposed architecture shown in Figure 1. |
| $\mathbb{C}$ | Set of IP/WDM core nodes, where $\mathbb{C} \subset \mathbb{N}$. |
| $\mathbb{A}$ | Set of Wi-Fi access points (APs), where $\mathbb{A} \subset \mathbb{N}$. |
| $\mathbb{O}$ | Set of ONU devices in the PON, where $\mathbb{O} \subset \mathbb{N}$. |
| $\mathbb{OT}$ | Set of OLT devices in the PON, where $\mathbb{OT} \subset \mathbb{N}$. |
| $\mathbb{MS}$ | Set of Ethernet switches in the metro, where $\mathbb{MS} \subset \mathbb{N}$. |
| $\mathbb{DC}$ | Set of cloud data centers (DCs), where $\mathbb{DC} \subset \mathbb{N}$. |
| $\mathbb{I}$ | Set of generic IoT devices, where $\mathbb{I} \subset \mathbb{N}$. |
| $\mathbb{P}$ | Set of nodes with processing capability, where $\mathbb{P} \subset \mathbb{N}$ and $\mathbb{P} = \mathbb{I} \cup \mathbb{O} \cup \mathbb{OT} \cup \mathbb{MS} \cup \mathbb{DC}$. |
| $\mathbb{S}$ | Set of all IoT devices generating task requests, where $\mathbb{S} \subset \mathbb{I}$. |

**IP/WDM Core Network Parameters:**

| | |
|---|---|
| $Pr$ | Maximum power consumption of an IP router port. |
| $Pt$ | Maximum power consumption of a transponder. |
| $Pe$ | Maximum power consumption of an EDFA. |
| $Po$ | Maximum power consumption of an optical switch. |
| $Prg$ | Maximum power consumption of a regenerator. |
| $Ir$ | Idle power consumption of an IP router port. |
| $It$ | Idle power consumption of a transponder. |
| $Ie$ | Idle power consumption of an EDFA. |
| $Io$ | Idle power consumption of an optical switch. |
| $Irg$ | Idle power consumption of a regenerator. |
| $B$ | Maximum data rate of a single wavelength. |
| $W$ | Number of wavelengths in a fibre. |
| $\epsilon^{(r)}$ | Energy per bit of a router port, where $\epsilon^{(t)} = \left(\frac{Pt - It}{B}\right)$. |
| $\epsilon^{(t)}$ | Energy per bit of a transponder, where $\epsilon^{(r)} = \left(\frac{Pr - Ir}{B}\right)$. |
| $\epsilon^{(e)}$ | Energy per bit of the EDFAs, where $\epsilon^{(e)} = \left(\frac{Pe - Ie}{B}\right)$. |
| $\epsilon^{(o)}$ | Energy per bit of the optical switches, where $\epsilon^{(o)} = \left(\frac{Po - Io}{B}\right)$. |
| $\epsilon^{(rg)}$ | Energy per bit of the regenerators, where $\epsilon^{(rg)} = \left(\frac{Prg - Irg}{B}\right)$. |
| $D_{mn}$ | Distance between core node $m$ and core node $n$, where $m, n \in C$. |
| $Se$ | Span distance between neighboring EDFAs. |
| $Sg$ | Span distance between two neighboring regenerators. |
| $A_{mn}$ | Number of EDFAs utilized on each fiber in the core network from node $m \in \mathbb{C}$ to $n \in \mathbb{C}$, where $A_{mn} = \left\lfloor \left(\frac{D_{mn}}{Se}\right) - 1 \right\rfloor + 2$. |
| $R_{mn}$ | Number of regenerators utilized between core node $m \in \mathbb{C}$ and core node $n \in \mathbb{C}$, $R_{mn} = \left\lfloor \left(\frac{D_{mn}}{Sg}\right) - 1 \right\rfloor$. |
| $PUE\_C$ | Power Usage Effectiveness of IP/WDM core network node. |

**Cloud Data Center Parameters:**

| | |
|---|---|
| $P^{(DS)}$ | Maximum power consumption of cloud DC switch. |
| $I^{(DS)}$ | Idle power consumption of cloud DC switch. |
| $B^{(DS)}$ | Data rate of cloud DC switch. |
| $\epsilon^{(DS)}$ | Cloud DC switch energy per bit, where $\epsilon^{(DS)} = \left(\frac{P^{(DS)} - I^{(DS)}}{B^{(DS)}}\right)$. |
| $P^{(DR)}$ | Maximum power consumption of cloud DC router. |
| $I^{(DR)}$ | Idle power consumption of cloud DC router. |
| $B^{DR}$ | Cloud DC router data rate. |
| $\epsilon^{(DR)}$ | Energy per bit of a Cloud DC router, where $\epsilon^{(DR)} = \left(\frac{P^{(DR)} - I^{(DR)}}{B^{(DR)}}\right)$. |

$PUE\_DC$    Power Usage Effectiveness of DC node, for processing and networking.

**Metro Network and Fog Parameters:**

$P^{(MS)}$    Maximum power consumption of a metro switch.

$I^{(MS)}$    Idle power consumption of a metro switch.

$B^{MS}$    Bit rate of a metro switch.

$\epsilon^{(MS)}$    Metro switch energy per bit, where $\epsilon^{(MS)} = \left(\frac{P^{(MS)} - I^{(MS)}}{B^{(MS)}}\right)$.

$P^{(MfS)}$    Maximum power consumption of a metro fog switch.

$I^{(MfS)}$    Idle power consumption of a metro fog switch.

$B^{MfS}$    Bit rate of a metro fog switch.

$\epsilon^{(MfS)}$    Metro fog switch energy per bit, where $\epsilon^{(MfS)} = \left(\frac{P^{(MfS)} - I^{(MfS)}}{B^{(MfS)}}\right)$.

$P^{(MR)}$    Maximum power consumption of a metro router.

$I^{(MR)}$    Idle power consumption of a metro router.

$B^{(MR)}$    Bit rate of a metro router.

$\epsilon^{(MR)}$    Metro router energy per bit, where $\epsilon^{(MR)} = \left(\frac{P^{(MR)} - I^{(MR)}}{B^{(MR)}}\right)$

$P^{(MfR)}$    Maximum power consumption of a metro fog router.

$I^{(MfR)}$    Idle power consumption of a metro fog router.

$B^{(MfR)}$    Bit rate of a metro fog router.

$\epsilon^{(MfR)}$    Metro fog router energy per bit, where $\epsilon^{(MfR)} = \left(\frac{P^{(MfR)} - I^{(MfR)}}{B^{(MfR)}}\right)$

$PUE\_M$    Power Usage Effectiveness of a metro node, for processing and networking.

$\mathcal{R}$    Metro router port redundancy.

**Access Network and Fog Parameters:**

$P^{(OT)}$    Maximum power consumption of OLT in the PON network.

$I^{(OT)}$    Idle power consumption of OLT in the PON network.

$B^{(OT)}$    Bit rate of OLT in the PON network.

$\epsilon^{(OT)}$    OLT router energy per bit, where $\epsilon^{(OT)} = \left(\frac{P^{(OT)} - I^{(OT)}}{B^{(OT)}}\right)$.

$P^{(O)}$    Maximum power consumption of an ONU in the PON network.

$I^{(O)}$    Idle power consumption of an ONU in the PON network.

$B^{(O)}$    Data rate of the Wi-Fi interface of an ONU device in the PON network.

$\epsilon^{(O)}$      ONU energy per bit, where $\epsilon^{(O)} = \left(\frac{P^{(O)} - I^{(O)}}{B^{(O)}}\right)$.

$P^{(AfS)}$      Maximum power consumption of an access fog switch.

$I^{(AfS)}$      Idle power consumption of an access fog switch.

$B^{AfS}$      Bit rate of an access fog switch.

$\epsilon^{(AfS)}$      Access fog switch energy per bit, where $\epsilon^{(AfS)} = \left(\frac{P^{(AfS)} - I^{(AfS)}}{B^{(AfS)}}\right)$.

$P^{(AfR)}$      Maximum power consumption of an access fog router.

$I^{(AfR)}$      Idle power consumption of an access fog router.

$B^{(AfR)}$      Bit rate of an access fog router.

$\epsilon^{(AfR)}$      Access fog router energy per bit, where $\epsilon^{(AfR)} = \left(\frac{P^{(AfR)} - I^{(AfR)}}{B^{(AfR)}}\right)$.

$P^{(CfR)}$      Maximum power consumption of CPE fog switch.

$I^{(CfR)}$      Idle power consumption of an CPE fog switch.

$B^{(CfR)}$      Bit rate of a CPE fog switch.

$\epsilon^{(CfR)}$      CPE fog switch energy per bit, where $\epsilon^{(CfR)} = \left(\frac{P^{(CfR)} - I^{(CfR)}}{B^{(CfR)}}\right)$.

$P^{(ap)}$      Maximum power consumption of an AP.

$I^{(ap)}$      Idle power consumption of an AP.

$B^{(ap)}$      Data rate of the AP.

$\epsilon^{(ap)}$      AP Wi-Fi interface energy per bit, where $\epsilon^{(ap)} = \left(\frac{P^{(ap)} - I^{(ap)}}{B^{(ap)}}\right)$.

$PUE\_A$      Power Usage Effectiveness of an access fog node, for processing and networking.

**Parameters of IoT Devices:**

$P^{(iot)}$      Maximum power consumption of an IoT transceiver.

$I^{(iot)}$      Idle power consumption of an IoT transceiver.

$B^{(iot)}$      Data rate of the Wi-Fi interface of an IoT device.

$\epsilon^{(iot)}$      IoT Wi-Fi interface energy per bit, where $\epsilon^{(iot)} = \left(\frac{P^{(iot)} - I^{(iot)}}{B^{(iot)}}\right)$.

**Parameters of Processing Devices:**

$P^{(cpu)}$      Maximum power consumption of processing device $d \in \mathbb{P}$, in Watts.

$I^{(cpu)}$      Idle power consumption of processing device $d \in \mathbb{P}$, in Watts.

| | |
|---|---|
| $C^{(cpu)}$ | Maximum capacity of processing device $d \in \mathbb{P}$ in Million Instructions Per Second (MIPS). |
| $E^{(mips)}$ | Energy per instruction of processing device $d \in \mathbb{P}$, where $E^{(mips)} = \left(\frac{P^{(cpu)} - I^{(cpu)}}{C^{(cpu)}}\right)$. |

**Application Parameters:**

| | |
|---|---|
| $D_s^{(cpu)}$ | Processing task in MIPS requested by source node $s \in \mathbb{S}$. |
| $T_s^{(cpu)}$ | Data rate traffic in Mbps requested by source node $s \in \mathbb{S}$. |
| $C_{mn}$ | Capacity of link $(m,n)$, where $m \in N$ and $n \in \mathbb{N}_m$. |
| $\delta$ | Portion of the idle power of equipment attributed to the use case. |
| $\Delta$ | Number of MIPS required to process 1Mb of traffic. |
| $M$ | Large enough number. |

**Variables:**

| | |
|---|---|
| $L^{sd}$ | Traffic demand between source node $s \in \mathbb{S}$ and processing device $d \in \mathbb{P}$. |
| $L_{mn}^{sd}$ | Traffic flow between source node $s \in \mathbb{S}$ and processing device $d \in \mathbb{P}$, traversing node $m \in \mathbb{N}$ and $n \in N_m$. |
| $L_d$ | Volume of aggregated traffic by node $d \in \mathbb{N}$. |
| $\mathcal{B}_m$ | $\mathcal{B}_m = 1$, if network node $m \in \mathbb{N}$ is activated, otherwise $\mathcal{B}_m = 0$. |
| $\theta_d$ | Traffic in node $d \in \mathbb{P}$ for processing, where $\theta_d = \lambda_d \Omega_d$. |
| $\Gamma_{mn}$ | If $\Gamma_{mn} = 1$, link $(m,n)$ in the core network, where $m \in \mathbb{C}, n \in (\mathbb{N}_m \cap \mathbb{C})$ is used, otherwise $\Gamma_{mn} = 0$. |
| $\rho^{sd}$ | Processing task of source node $s \in \mathbb{S}$ allocated to processing device $d \in \mathbb{P}$. |
| $\Omega^{sd}$ | $\Omega^{sd} = 1$, if processing task of source node $s \in \mathbb{S}$ is allocated to destination node $d \in \mathbb{P}$, otherwise $\Omega^{sd} = 0$. |
| $\Omega^d$ | $\Omega^d = 1$, if processing node $d \in \mathbb{P}$ is turned ON, otherwise $\Omega^d = 0$. |
| $\mathcal{N}_d$ | Number of processing servers used at node $d \in \mathbb{P}$. |
| $W_{mn}$ | Number of wavelengths used in fiber link $(m,n)$, where nodes $m, n \in \mathbb{C}$. |
| $F_{mn}$ | Number of fibers used on link $(m,n) \in \mathbb{C}$. |
| $Ag_m$ | Number of core router aggregation ports activated at node $m \in \mathbb{C}$. |

In this chapter, we adopt the power profile depicted in **Figure 2**, which consists of an idle and proportional section. The idle section is consumed as soon as the device is turned ON, regardless of the load (MIPS or traffic). Whereas the proportional section is dependent on the amount of workload (processing and /or traffic) that is allocated to the device. Almost all devices adopt a linear power profile similar to the one shown in **Figure 2** [47]. Hence, in practical settings, idle power represents a large proportion of the maximum power of a device (networking or processing) and therefore cannot be ignored.

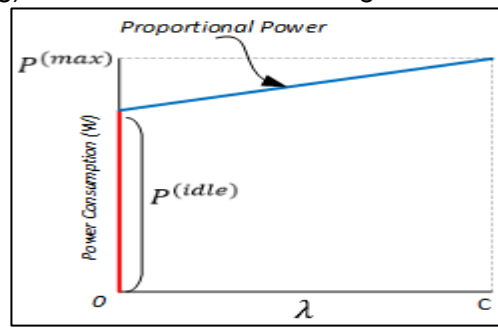

**Figure 2** The adopted power profile with two parts; a) proportional power consumption and b) idle power consumption.

The architecture considered spans across multiple layers of processing and networking. Therefore, it becomes a necessity to fairly represent the utilization characteristics of both the networking and processing devices. In the literature, when idle, servers are reported to consume around 60% of their maximum power consumption, whilst networking nodes are reported to consume around 90% of their maximum power consumption [48]. In this chapter, both ratios are assumed for the idle power consumption of both networking and processing elements. However, large networking equipment, such as those found in the access to the core layer, are assumed to consume a portion (3% based on [49]) of the total idle power consumption. This is a reasonable assumption since such devices can be shared by many applications due to the number of users that can be connected to them. The total power consumption given the power profile in **Figure 2** is calculated using equation (1):

$$Total\ Power\ Consumption = \left(\frac{P^{(\max)} - P^{(idle)}}{C}\right)\lambda + P^{(idle)} \qquad (1)$$

where $P^{(max)}$ is the maximum power consumption of the device (networking or processing) which is consumed as soon as the device is activated regardless of the load $\lambda$ and ($Pmax$) is the maximum power consumption of the device, when it is utilized at full capacity C. The linear curve represents the proportional power consumption. For networking devices, this is expressed as energy per bit and likewise, for processing, it is expressed as energy per instruction.

The total power consumption of the proposed cloud fog architecture is composed of the power consumption in processing nodes and the power consumption in the network. The processing power consumption also includes the power consumption of the networking elements needed for intra processing node's communication.

**1) Network Power Consumption** ($net\_pc$):

The total power consumption in the core network, under the non-bypass light path approach [50] is composed of:

The power consumption of core router ports:

$$PUE\_C \left[ \sum_{m \in \mathbb{C}} \left( \epsilon^{(r)} L_m \right) + \sum_{m \in \mathbb{C}} \left( \delta Ir \left( Ag_m + \sum_{n \in (\mathbb{N}_m \cap \mathbb{C})} W_{mn} \right) \right) \right] \quad (2)$$

The power consumption of transponders:

$$PUE\_C \left[ \sum_{m \in \mathbb{C}} \left( \epsilon^{(t)} L_m \right) + \sum_{m \in \mathbb{C}} \sum_{n \in (\mathbb{N}_m \cap \mathbb{C})} (\delta It W_{mn}) \right] \quad (3)$$

The power consumption of EDFAs:

$$PUE\_C \left[ \sum_{m \in \mathbb{C}} \left( \epsilon^{(t)} L_m A_{mn} F_{mn} \right) + \sum_{m \in \mathbb{C}} \sum_{n \in (\mathbb{N}_m \cap \mathbb{C})} (\delta Ie A_{mn} F_{mn}) \right] \quad (4)$$

The power consumption of optical switches:

$$PUE\_C \left[ \sum_{m \in \mathbb{C}} \left( \epsilon^{(o)} L_m \right) + \sum_{m \in \mathbb{C}} (\delta Io \mathcal{B}_m) \right] \quad (5)$$

The power consumption of regenerators:

$$PUE\_C \left[ \sum_{m \in \mathbb{C}} \left( \epsilon^{(rg)} L_m Rg_{mn} W_{mn} \right) + \sum_{m \in \mathbb{C}} \sum_{n \in (\mathbb{N}_m \cap \mathbb{C})} (Irg \, Rg_{mn} W_{mn}) \right] \quad (6)$$

The metro network power consumption consists of the power consumption of metro routers and switches, which is given as:

$$PUE\_M \left[ \mathcal{R} \sum_{m \in \mathbb{MR}} \left( \epsilon^{(MR)} L_m \right) + \mathcal{R} \sum_{m \in \mathbb{MR}} \left( \delta I^{(MR)} \mathcal{B}_m \right) + \sum_{m \in \mathbb{MS}} \left( \epsilon^{(MS)} L_m \right) \right. \\ \left. + \sum_{m \in \mathbb{MS}} \left( \delta I^{(MS)} \mathcal{B}_m \right) \right] \quad (7)$$

The power consumption of the PON access network comprises of the power consumption of OLT and ONU devices:

$$PUE\_A \left[ \sum_{m \in \mathbb{OT}} \left( \epsilon^{(OT)} L_m \right) + \sum_{m \in \mathbb{OT}} \left( \delta I^{(OT)} \mathcal{B}_m \right) + \sum_{m \in \mathbb{O}} \left( \epsilon^{(\mathbb{O})} L_m \right) + \sum_{m \in \mathbb{O}} \left( \delta I^{(O)} \mathcal{B}_m \right) \right] \quad (8)$$

The Wi-Fi AP's power consumption is given as:

$$\sum_{m\in\mathbb{AP}} \left(\epsilon^{(ap)} L_m\right) + \sum_{m\in\mathbb{AP}} \left(\delta I^{(ap)} \mathcal{B}_m\right) \quad (9)$$

The IoT devices' transceivers power consumption is given as:

$$\sum_{m\in\mathbb{I}} \left(\epsilon^{(iot)} L_m\right) + \sum_{m\in\mathbb{I}} \left(\delta I^{(iot)} \mathcal{B}_m\right) \quad (10)$$

**2) Processing Power Consumption ($pr\_pc$):**

The total power consumption of the processing nodes is composed of:

The processing power consumption of IoT devices:

$$\sum_{s\in\mathbb{S}} \sum_{d\in\mathbb{I}} \left(E_d^{(i)} \rho^{sd}\right) + \sum_{d\in\mathbb{I}} \left(I^{(pr)} \mathcal{N}_d\right) \quad (11)$$

The processing power consumption of CPE fog servers:

$$\sum_{s\in\mathbb{S}} \sum_{d\in\mathbb{O}} \left(E_d^{(i)} \rho^{sd}\right) + \sum_{d\in\mathbb{O}} \left(I_d^{(pr)} \mathcal{N}_d\right) \quad (12)$$

The processing power consumption of access fog servers:

$$\mathbb{P}a \left[\sum_{s\in\mathbb{S}} \sum_{d\in\mathbb{OT}} \left(E_d^{(i)} \rho^{sd}\right) + \sum_{d\in\mathbb{OT}} I_d^{(pr)} \mathcal{N}_d\right] \quad (13)$$

The processing power consumption of metro fog servers:

$$\mathbb{P}m \left[\sum_{s\in\mathbb{S}} \sum_{d\in\mathbb{MS}} \left(E_d^{(i)} \rho^{sd}\right) + \sum_{d\in\mathbb{MS}} \left(I_d^{(pr)} \mathcal{N}_d\right)\right] \quad (14)$$

The processing power consumption of cloud data center (DC) servers:

$$\mathbb{P}d \left[\sum_{s\in\mathbb{S}} \sum_{d\in\mathbb{DC}} \left(E_d^{(i)} \rho^{sd}\right) + \sum_{d\in\mathbb{DC}} \left(I_d^{(pr)} \mathcal{N}_d\right)\right] \quad (15)$$

The intra cloud DC power consumption is composed of the power consumption of the cloud LAN, which consist of a router and a switch:

$$\mathbb{P}d \left[\sum_{d\in\mathbb{DC}} \left(\epsilon^{(DR)} \theta_d\right) + \sum_{d\in\mathbb{DC}} \left(\delta I^{(DR)} \Omega^d\right) + \sum_{d\in\mathbb{DC}} \left(\epsilon^{(DS)} \theta_d\right) + \sum_{d\in\mathbb{DC}} \left(\delta I^{(DS)} \Omega^d\right)\right] \quad (16)$$

The intra metro fog power consumption consists of power consumption of metro fog routers and switches:

$$\mathbb{P}m \left[ \sum_{d \in \mathbb{MS}} \left( \epsilon^{(MfR)} \theta_d \right) + \sum_{d \in \mathbb{MS}} \left( \delta \epsilon^{(MfR)} \Omega^d \right) + \sum_{d \in \mathbb{MS}} \left( \epsilon^{(MfS)} \theta_d \right) + \sum_{d \in \mathbb{MS}} \left( \delta I^{(MfS)} \Omega^d \right) \right] \quad (17)$$

The MILP model's objective is to minimize the total power consumption as follows:

**Minimize**: $net_{pc} + pr\_pc$

**Subject to the following constraints:**

$$\sum_{n \in \mathbb{N}_m} L_{mn}^{sd} - \sum_{n \in \mathbb{N}_m} L_{nm}^{sd} = \begin{cases} L_{sd} & m = s \\ -L_{sd} & m = d \\ 0 & otherwise \end{cases} \quad \forall s \in \mathbb{S}, d \in \mathbb{P}, m \in \mathbb{N}: s \neq d. \quad (18)$$

Constraint (18) is the flow conservation constraint. It ensures that the total incoming traffic at a node is equal to the total outgoing traffic of that node; if the node is not a source or a destination node.

$$\sum_{d \in \mathbb{P}} \rho^{sd} = D_s^{(cpu)} \quad \forall s \in \mathbb{S} \quad (19)$$

Constraint (19) ensures that processing service demand per IoT source node $s \in S$ is met at a given destination node.

$$\rho^{sd} \geq \Omega^{sd} \quad \forall s \in \mathbb{S}, d \in \mathbb{P} \quad (20)$$

$$\rho^{sd} \leq M\Omega^{sd} \quad \forall s \in \mathbb{S}, d \in \mathbb{P} \quad (21)$$

Constraints (20) and (21) are used to ensure that the binary variable $\rho^{sd} = 1$ if destination node $d \in P$ is activated to host the processing demand of source node $s \in S$.

$$\sum_{d \in \mathbb{P}} \Omega^{sd} \leq K \quad \forall s \in \mathbb{S} \quad (22)$$

Constraint (22) ensures that the number of sub-services a processing demand can be divided into is less than or equal to K, hence $K = 1$ implies no service splitting is allowed.

$$\mathcal{N}_d \leq \mathcal{V}_d \quad \forall d \in \mathbb{P} \quad (23)$$

Constraint (23) ensures that the number of servers activated at a processing node $d \in P$, does not exceed the maximum available number of servers in that node.

$$\sum_{s \in \mathbb{I}} \Omega^{sd} \geq \Omega^d \quad \forall d \in \mathbb{P} \tag{24}$$

$$\sum_{s \in \mathbb{I}} \Omega^{sd} \leq M \Omega^d \quad \forall d \in \mathbb{P} \tag{25}$$

Constraints (24) and (25) are used to ensure that, the binary variable $\Omega^d = 1$ if processing node $d \in \mathbb{P}$ is activated, otherwise $\Omega^d = 0$.

$$\lambda_m = \sum_{\substack{s \in \mathbb{S}: \\ m=s}} \sum_{d \in \mathbb{P}} \sum_{n \in \mathbb{N}_m} L^{sd}_{mn} + \sum_{\substack{s \in \mathbb{S}: \\ m \neq s}} \sum_{\substack{d \in \mathbb{P}: \\ s \neq d}} \sum_{n \in \mathbb{N}_m} L^{sd}_{nm} \quad \forall m \in S \tag{26}$$

$$L_m = \sum_{\substack{s \in \mathbb{S}: \\ m \neq s}} \sum_{\substack{d \in \mathbb{P}: \\ s \neq d}} \sum_{n \in \mathbb{N}_m} L^{sd}_{nm} \quad \forall m \in (\mathbb{I} \cup \mathbb{AP} \cup \mathbb{O} \cup \mathbb{OT} \cup \mathbb{MS} \cup \mathbb{MR} \cup \mathbb{DC}) \tag{27}$$

$$L_m = \sum_{s \in \mathbb{S}} \sum_{\substack{d \in \mathbb{P}: \\ s \neq d}} \sum_{\substack{n \in \mathbb{N}_m: \\ n \in (\mathbb{N}_m \cap \mathbb{C})}} L^{sd}_{mn} \quad \forall m \in \mathbb{C} \tag{28}$$

Constraint (26) gives the traffic generated or received by an IoT node with the first term representing its role as a source and the second term representing IoT node serving demands of other IoT nodes. Constraint (27) gives the traffic traversing / received by a node of the access, metro, and cloud network. Constraint (28) gives the traffic traversing the core nodes.

$$\theta_d \leq M\Omega^d \quad \forall d \in \mathbb{P} \tag{29}$$

$$\theta_d \leq L_d \quad \forall d \in \mathbb{P} \tag{30}$$

$$\theta_d \geq \lambda_d - (1 - \Omega^d)M \quad \forall d \in \mathbb{P} \tag{31}$$

Constraints (29), (30) and (31) are used to linearize the non-linear equation $\lambda_d \Omega_d$, where $d \in P$. This ensures that traffic on a processing node $d \in P$ is only accounted for if it is destined to that node for processing.

$$L_m \geq \mathcal{B}_m \quad \forall m \in \mathbb{N} \tag{32}$$

$$L_m \leq M\mathcal{B}_m \quad \forall m \in \mathbb{N} \tag{33}$$

Constraints (32) and (33) are used to ensure that, the binary variable $\mathcal{B}_m = 1$ if network node $m \in N$ is activated, otherwise $\mathcal{B}_m = 0$.

$$L^{sd} = T^{(DR)} \Omega_{sd} \quad \forall s \in \mathbb{S}, d \in \mathbb{P} \tag{34}$$

Constraint (34) ensures that traffic is only directed to the destination node that is hosting a processing service.

$$\sum_{s \in \mathbb{S}} \sum_{\substack{d \in \mathbb{P}: \\ s \neq d}} L^{sd}_{mn} \leq C_{mn} \quad \forall m \in (\mathbb{I} \cup \mathbb{AP} \cup \mathbb{O} \cup \mathbb{OT} \cup \mathbb{MS} \cup \mathbb{MR} \cup \mathbb{DC}): n \in \mathbb{N}_m \tag{35}$$

Constraint *(35)* ensures that the total traffic carried on link $m, n$, in all layers except the core does not exceed the link capacity.

$$Ag_m \geq \frac{L_m}{B} \quad \forall m \in \mathbb{C} \tag{36}$$

Constraint (36) gives the number of aggregation router ports at each IP/WDM node.

$$\sum_{s \in \mathbb{S}} \sum_{\substack{d \in \mathbb{P}: \\ s \neq d}} L_{mn}^{sd} \leq W_{mn} B \quad \forall m \in \mathbb{C}: n \in (\mathbb{C} \cap \mathbb{N}_m) \tag{37}$$

$$W_{mn} \leq WF_{mn} \quad \forall m \in \mathbb{C}: n \in (\mathbb{C} \cap \mathbb{N}_m) \tag{38}$$

Constraints (37) and (38) represent the physical link capacity of the IP/WDM optical links. Constraint (37) ensures that the total traffic on a link does not exceed the capacity of a single wavelength while constraint (38) ensures the total number of wavelength channels does not exceed the capacity of a single fiber link.

# Input Data for the MILP Model

## Processing and Data Rates

In this chapter, we assume that processing task requirement is proportional to data rate, such that, for every bit of traffic, 1000 MIPS is required for processing. This assumption is based on the work in [51] where for a file of 10 kB, 69.23 MIPS are required for processing for visual processing applications. Thus, through we derive how many MIPS are required (Δ) to process a Mb of traffic using (39):

$$\Delta = \frac{69.23}{0.08} \cong 865.4. \tag{39}$$

For the sake of simplicity, we assume that a Mb of traffic requires approximately 1000 MIPS for processing. As for the bandwidth requirement, an online tool is used to estimate the required data rates for different video resolutions and this was estimated to be between 1 – 10 Mbps, which covers video resolutions between $1024 \times 720$ to $1600 \times 1200$ at 30 frames per second [52]. The CPU workload intensity is then calculated by multiplying the Δ by the amount of traffic. Thus, this makes the CPU demand proportional to the size of the traffic due to the assumption that the higher the traffic, the more features a video file will hold.

## Power Consumption Data

The data for all the network devices in the network (except the core as it is shown in separate table) is shown in Table 1. We have use of manufacturer's and equipment datasheet where possible in order to represent a practical scenario. The idle power consumption of high capacity networking equipment is reported to consume up to 90% of equipment's maximum power consumption [40]. Since high capacity networking equipment is shared by many users and applications, we assume that the IoT application under consideration only consumes 3% ($\delta$) of the equipments' maximum idle power consumption. This is based on Cisco's visual networking index for the years 2017-2022. It is reported that, globally, 3% of all video traffic on the Internet is due to surveillance applications [49]. As for processing device's idle power consumption, based on [48], we assume it is 60% of the maximum power consumption of the CPU. The processing devices' input data are summarized in Table 2. We estimate the processing capacity of processing nodes (in MIPS) using a technical benchmark published in

[53]. It is reported that high-end CPUs process 4 instructions per cycle (I/C). Thus, to determine the maximum capacity of a processing device we have used the following equation

$$\text{MIPS} = \text{clock} \times \text{I/C} \qquad (40)$$

where $I/C$ is the number of instructions a CPU can execute per clock cycle in GHz. To differentiate between the different types of CPUs and their efficiencies, we set the $I/C$ of the Metro Fog server as a reference point. The efficiency of the processing decreases as one moves down the network hierarchy (from the core to the edge) [54]. At those layers where multiple servers can be deployed, networking infrastructure becomes a necessity to establish a LAN network between multiple active servers. Hence, we have used routers and switches accordingly to achieve this and Table 3 contains the data of all the devices utilized for this purpose. For the lower layers of the cloud fog architecture such as IoT and CPE, we assume embedded type processors such as Raspberry Pi (RPi) Zero W and Raspberry Pi (RPi) 3 Model B, respectively. We assume the cloud DC node is a single hop away from the aggregated traffic and the average distance is also assumed to span 2010 km on average, which is estimated using google maps for AT&T US network topology [55]. The power consumption of the IP/WDM core network is consistent with our previous work in [13] and all the parameters are summarized in Table 4.

| Node | Maximum Power (W) | Idle Power (W) | $\delta$ | Data Rate (Gb/s) |
|---|---|---|---|---|
| IoT (WiFi) | 0.56 [56] | 0.34 [57] | - | 0.1 [56] |
| ONU (WiFi) | 15 [58] | 9 [58] | - | 0.3 [58] |
| OLT | 1940 [59] | 60 [59] | 3% | 8600 [59] |
| Metro Router Port | 30 [60] | 27 | 3% | 40 [60] |
| Metro Ethernet Switch | 470 [61] | 423 | 3% | 600 [61] |
| Metro Router Redundancy (R) | 2 [41] | | | |

Table 1 Data for all networking devices in the network except the core layer

| Node | Device | Maximum Power (W) | Idle Power (W) | GHz | k MIPS | Watts /MIPS | Instruction Per Cycle |
|---|---|---|---|---|---|---|---|
| GP-DC Server | Intel Xeon E5-2680 | 130 [62] | 78 | 2.7 [62] | 108 | $481\mu$ | 5 |
| Metro Server | Intel X5675 | 95 [63] | 57 | 3.06 [63] | 73.44 | $517\mu$ | 4 |
| Access Server | Intel Xeon E5-2420 | 95 [64] | 57 | 1.9 [64] | 34.2 | $1111\mu$ | 3 |
| CPE Server | RPi 3 Model B | 12.5 [65] | 2 | 1.2 [66] | 2.4 | $4375\mu$ | 2 |
| IoT Server | RPi Zero W | 3.96 [65] | 0.5 | 1 [67] | 1 | $3460\mu$ | 1 |

Table 2 Data of all processing devices.

| Device | Maximum Power (W) | Idle Power (W) | Data Rate (Gb/s) | Energy Per Bit (W/Gb/s) |
|---|---|---|---|---|
| Access Fog Router | 13W[60] | 11.7 | 40[60] | 0.03 |
| Access Fog Switch | 210W[61] | 189 | 240[61] | 0.08 |
| Metro Fog Router | 13W[60] | 11.7 | 40[60] | 0.03 |
| Metro Fog Switch | 210W [61] | 189 | 600[61] | 0.04 |
| DC LAN Router | 30[60] | 27 | 40[60] | 0.08 |
| DC LAN Switch | 470[61] | 423 | 600[61] | 0.08 |

Table 3 Data for networking devices used inside (intra-processing) Access Fog, Metro Fog and Data Centre processing units.

| | |
|---|---|
| Distance between two neighbouring EDFAs $Se$ | 80 (km) [13] |
| Number of wavelengths in a fibre ($W$) | 32 [13] |
| Bitrate of a wavelength ($B$) | 40 Gb/s |
| Distance between two neighbouring core nodes $D_{mn}$ | 2500km |
| Maximum power consumption of a router port $Pr$ | 638 (W) [13] |
| Idle power consumption of a router port $Ir$ | 574.2 (W) |
| Energy per bit of a router port $\epsilon^{(r)}$ | 1.6 W/Gb/s |
| Maximum power consumption of a transponder $Pt$ | 129 (W) [13] |
| Idle power consumption of a transponder $It$ | 116 (W) |
| Energy per bit of a transponder $\epsilon^{(t)}$ | 0.32 (W/Gb/s) |
| Maximum power consumption of an optical switch $Po$ | 85 (W) [13] |
| Idle power consumption of an optical switch $Io$ | 76.5 (W) |
| Energy per bit of an optical switch $\epsilon^{(o)}$ | 0.2 (W/Gb/s) |
| Maximum power consumption of a regenerator that reaches 2500km $Prg$ | 71.4 (W) [13] |
| Idle power consumption of a regenerator $Irg$ | 64 (W) |
| Energy per bit of a regenerator $\epsilon^{(rg)}$ | 0.19 (W/Gb/s) |

Table 4 Data for IP/WDM core network.

## Power Usage Effectiveness (PUE)

The power usage effectiveness (PUE) is a ratio that is used to measure the efficiency of a facility such as DCs, ISP networks, etc. PUE is defined as the ratio of the total power consumed by a facility to the total power consumed by the communications and processing elements within the facility. In DCs, Google reported that one of its DC has a PUE of 1.15 in 2018. In this chapter, we estimate the value of PUE on "space type", such that the value of PUE decreases with the increase in "space" [94]. Similarly, we increase PUE progressively in the proposed network architecture since the largest "Space Type" is generally occupied by cloud DCs connected to the core network. We assume that at the access and metro layers, processing and networking equipment have the same PUE as these two types of elements can be collocated in the same office/ building. The PUE value of the core network is consistent with one of our previous works, which is 1.5 [21]. Table 5 is a summary of the PUE values used in this chapter.

| Node | PUE |
|---|---|
| IoT | 1 |
| CPE | 1 |
| Access Fog ($PUE\_A$) | 1.5 |
| Metro Fog ($PUE\_M$) | 1.4 |
| Cloud DC ($PUE\_DC$) | 1.12 [68] |
| IP/WDM Core ($PUE\_C$) | 1.5 [21] |

Table 5 PUE values used in the MILP model.

# Scenarios and Processing Placement Results

## Energy-Aware Processing Placement

In this subsection, we consider a capacitated case where extra processing capacity cannot be added to the processing nodes in the cloud fog architecture. Such design problems are faced in the short-term when the network is already designed, and the processing nodes have been put in place. It is important to note that the cloud DC offers unlimited processing capacity and hence always enough to host all the tasks. We evaluate the performance of the proposed cloud fog architecture given a processing placement problem, and 20 generic IoT devices that are divided into 4 groups uniformly. The total number of IoT devices in each group is based on a representative home LAN network which typically connects a single to few users [69]. The performance of the cloud fog approach is benchmarked against the baseline approach in which all of the task requests are simultaneously processed by the cloud DC and the types of scenarios evaluated in this sub-section are shown in Table 6.

| Use Case | Source Node Distribution | Total # of Source nodes | Total Requested MIPS | |
|---|---|---|---|---|
| | | | Min | Max |
| Scenario #One | A single task request generated in any random IoT group. | 1 | 1k | 10k |
| Scenario #Two | Five task requests generated in the same IoT group. | 5 | 5k | 100k |
| Scenario #Three | Four task requests generated per IoT group. | 4 | 4k | 80k |
| Scenario #Four | Five task requests generated per IoT group. | 20 | 20k | 200k |

Table 6 Types of scenarios evaluated in this sub-section.

## Scenario One

In this scenario, it is assumed that at any given time, a single source node is generating task requests from any IoT group randomly. Figure 3 shows the total power consumption of the cloud fog approach versus the baseline solution. As was expected, during low workloads such as at 1000 MIPS, processing tasks are allocated to source nodes themselves and when processing capacity runs out at this layer, the model utilizes the closest processing layer to the source nodes, which is the CPE layer. This is justifiable due to the low-power embedded type CPUs onboard these devices and their close proximity to the source nodes, hence substantially lower networking and processing power consumption as can be seen in Figure 4(a). However, once the processing resources run out at the two aforementioned layers, the model utilizes the large servers located at the Access Fog. As shown in Figure 3, allocating processing tasks to nodes with low-power embedded CPUs introduces substantial power savings of up to 98% and up to 46% when tasks are processed by larger fog servers. The allocation of the total processing tasks (in %) in the different processing layers is shown in Figure 5. Although this result is optimal in this situation, it may not be so optimal when the processing capacity of the CPE layer can be expanded (i.e. evaluations done in an un-capacitated setting). The reason behind this is that CPE nodes consume considerably lower power compared with the servers at the Access Fog and they do not have any associated overheads such as PUE due to cooling requirements. Also, it is worthy of noting that the location of the source in the network highly influences the allocation of the processing tasks due to the power consumption of the network in order to access certain processing layer. It is for this reason we evaluate different scenarios in which source nodes are in different parts of the network. Also, the reason why the baseline curve is flat is because the DC's processing efficiency is substantially higher than the rest of the processing locations, hence the proportional power consumption increases in very small steps. We would see a staircase curve should the number of servers increase due to the idle power consumption [35].

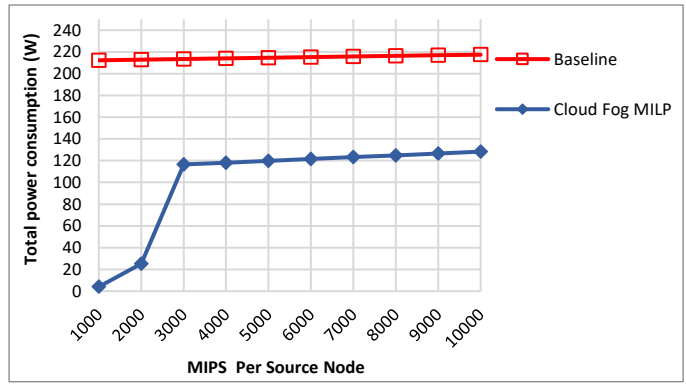

**Figure 3** Total power consumption in Scenario one of the cloud fog approach versus the baseline approach.

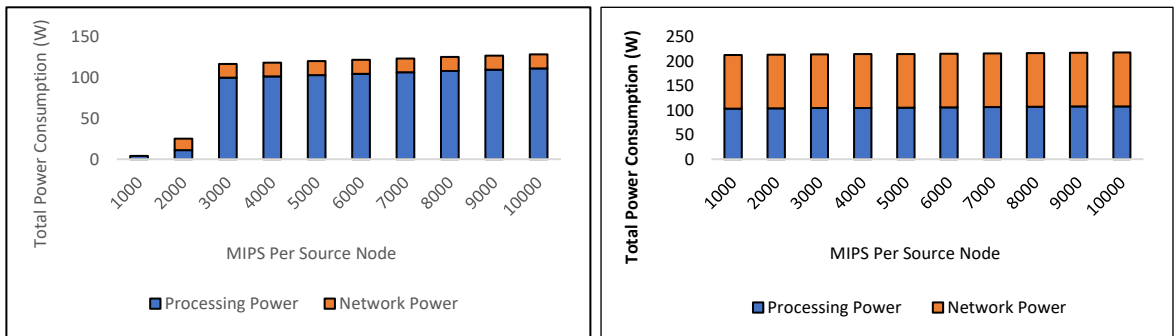

(a)                                                                 (b)

**Figure 4** Total power consumption in Scenario one broken down into network and processing power consumption in a) cloud fog approach and b) baseline approach.

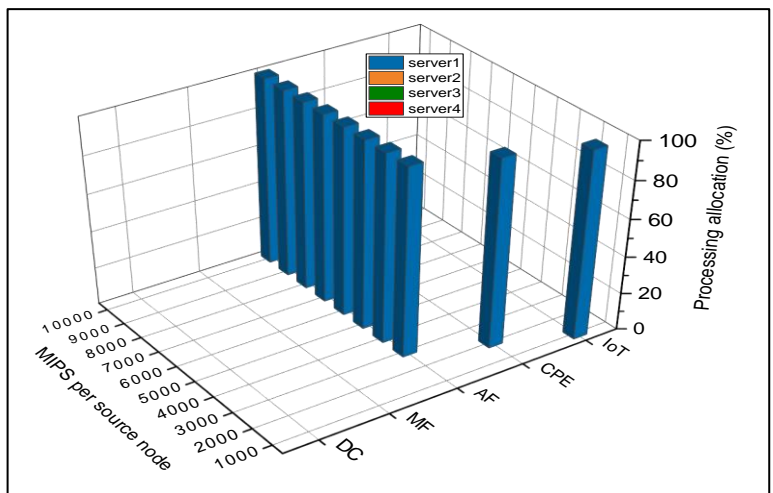

**Figure 5** Processing task allocation

## Scenario Two

In this scenario, we begin to observe the utilization of the Metro Fog node instead; and the Access Fog node has limited role to play in this case. This was anticipated because the Access Fog has a lower processing efficiency and a higher PUE value compared with the Metro Fog node. As shown in Figure 7(a), the Access Fog node is chosen to process the tasks at 2000 MIPS only because the network power consumption to access the Metro Fog node overrides the processing efficiency and lower PUE advantage of the Metro Fog node. However, as the workload increases (at and beyond 3000 MIPS), we can observe that the Metro Fog's efficiency compensates for its networking overhead, hence all tasks are processed at this layer. In Figure 6, the cloud fog approach still produces substantial power savings despite the activation of larger fog servers to process the increased demands such as those in the Access Fog and Metro Fog layers. At a workload of 2000 MIPS, we can observe the impact of the single allocation constraint (i.e. task request per source node cannot be split) on the total power savings. Although the CPE nodes had enough capacity to host the majority of the tasks (9600 MIPS out of 10,000 MIPS), the optimization model is forced to allocate all the tasks to a larger server with sufficient capacity such as the Access Fog node in this case. Had the model considered the prospect of task splitting and /or adding further processing capabilities to the CPE layer, the results would have been different. These dimensions are thoroughly investigated in our previous work in [70]. Task splitting can help realize server utilization improvements and can help better pack the lower processing and networking layers of the cloud fog architecture as we know from the previous scenario that the IoT and CPE layers produce substantial savings.

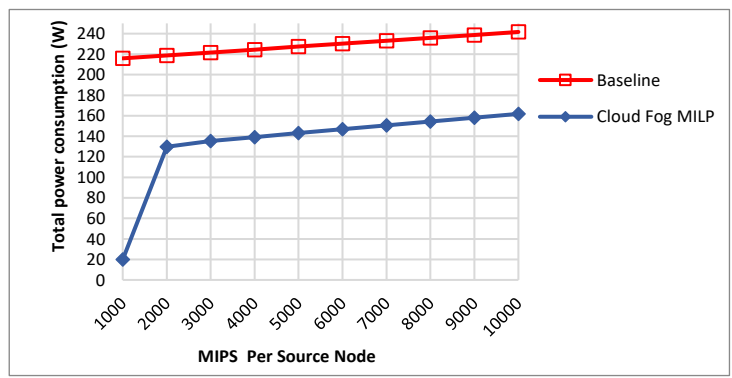

**Figure 6** Total power consumption in Scenario two of the cloud fog approach versus the baseline approach.

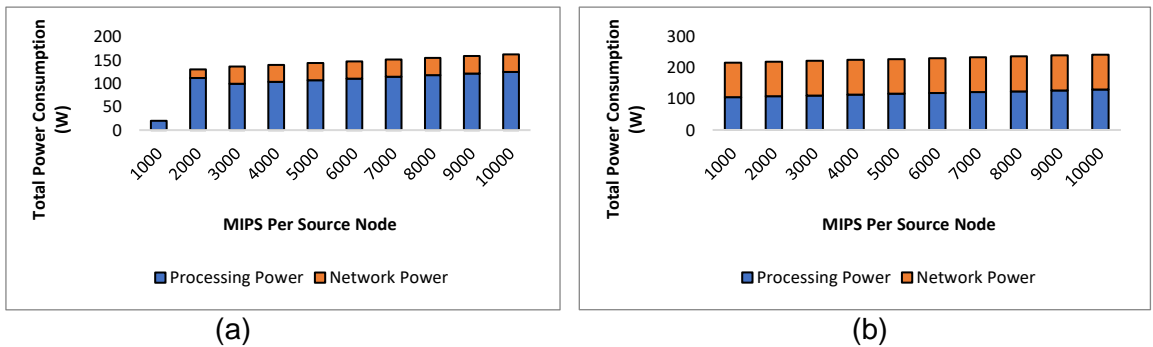

(a)                        (b)

**Figure 7** Total power consumption in Scenario two broken down into network and processing power consumption in a) cloud fog approach and b) baseline approach.

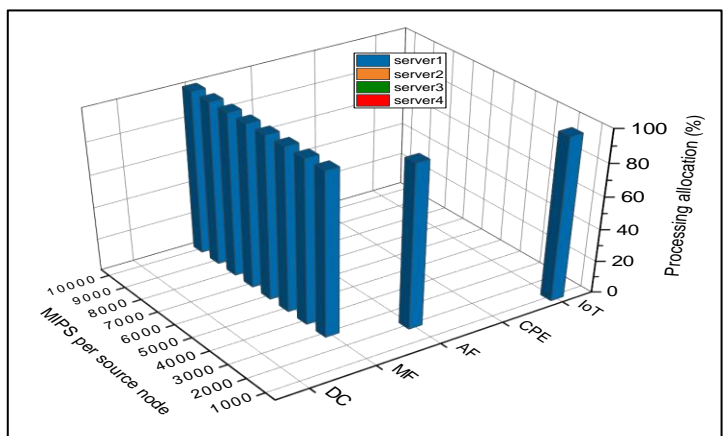

Figure 8 Processing task allocation

## Scenario Three

As shown in Figure 12, the trends in this scenario are similar to those observed in Scenario Two, except for the case at 2000 MIPS where instead of the Access Fog server, the model allocates the total demands to all of nodes in the CPE layer. This is primarily due to the geographical distribution of the source nodes in this scenario. Each CPE Fog server has enough processing capacity to process the task of the closest source node and the total demands happen to match the total processing capacity offered by the CPE layer. Hence, the model activates multiple low-power CPE servers in order to avoid the high idle power and associated PUE overheads of the higher fog layers such as the Access Fog and the Metro Fog nodes, as can be seen in Figure 10(a). A total power saving of 66% is achieved at 2000 MIPS and up to 55% power savings at workloads beyond 2000 MIPS, as shown in Figure 9.

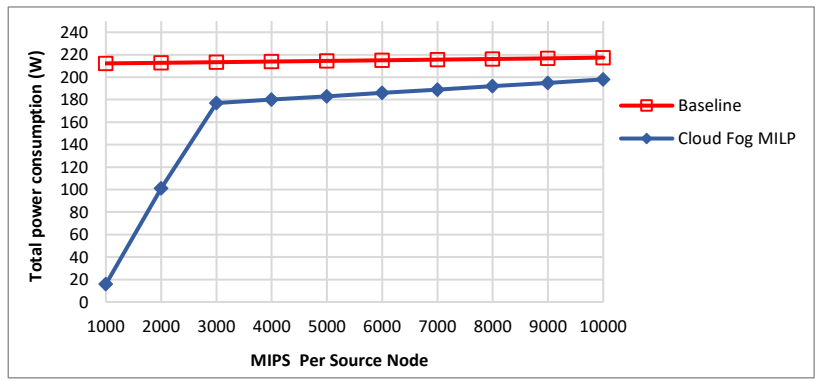

**Figure 9** Total power consumption in Scenario Three of the cloud fog approach versus the baseline approach.

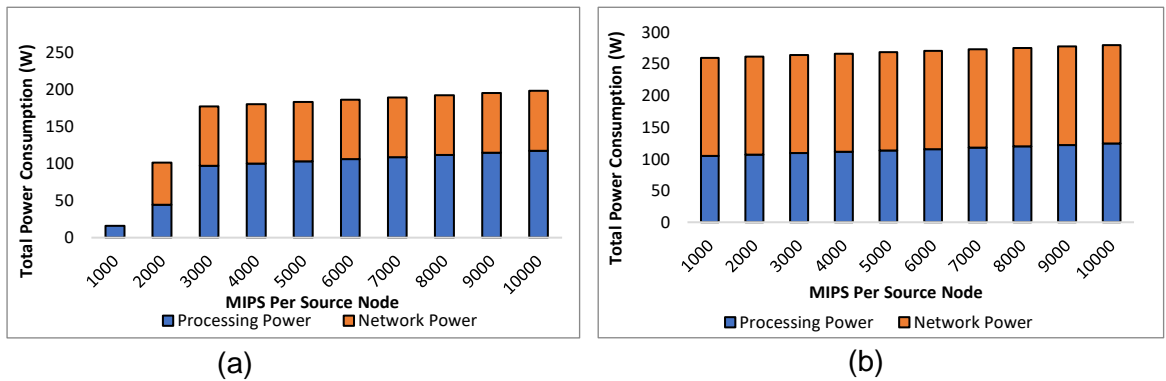

(a)                                    (b)

**Figure 10** Total power consumption in Scenario three broken down into network and processing power consumption in a) cloud fog approach and b) baseline approach.

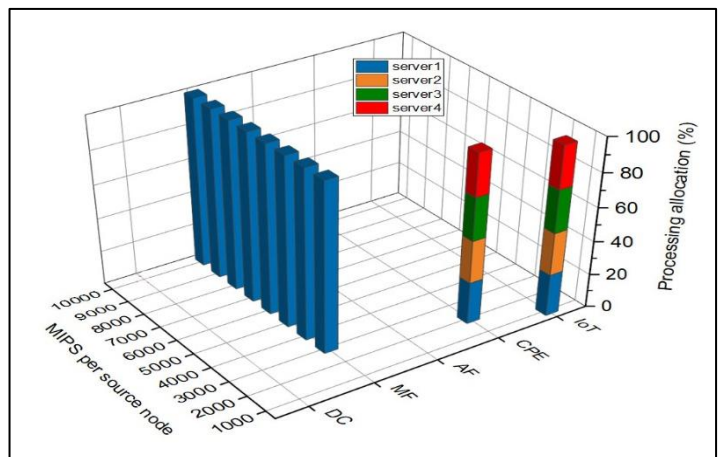

**Figure 11** Processing task allocation

## Scenario Four

In this scenario, all the of the source nodes generate task requests, hence the total workload volume has increased substantially. We begin to observe similar trends as in previous scenarios. At very low load workloads, processing locally on source nodes is still the optimal choice in terms of total energy consumption as shown in Figure 12. The allocation decisions shown in Figure 14 confirm the superiority of the Metro Fog server over the Access Fog server for high workloads, as the Access Fog is never utilized. The model chooses to utilize the Metro Fog layer at four different workloads due to the processing and networking trade-off shown in Figure 13(a) and Figure 13(b). At 4000 MIPS and 5000 MIPS, processing all the tasks in the Metro Fog layer results in activating 2 servers, therefore trading off the high network power consumption from accessing the cloud DC produces more power savings than choosing the Metro Fog. This is because the cloud DC has enough processing resources to server all the tasks on a single server, hence resulting in lower total server idle power consumption compared to the Metro Fog layer. However, at 6000 MIPS and 7000 MIPS, processing all the requests in the cloud DC results in the activation of two servers, hence the idle power consumption coupled with the core network power consumption renders the cloud DC solution no longer favourable and as a result the Metro Fog is chosen as the optimal location for processing all the tasks. At 8000 MIPS and beyond, the cloud DC solution due to its superiority in terms of processing capacity produces more power savings than the Metro Fog layer, hence it is chosen as the optimal allocation. Figure 12 shows power savings of up to 29% with the cloud fog approach. In this scenario, we show that the cloud fog approach does not replace the cloud DC, but instead the complementary features of both of these paradigms helps to achieve a better and more greener processing platform in the upcoming 6G networks.

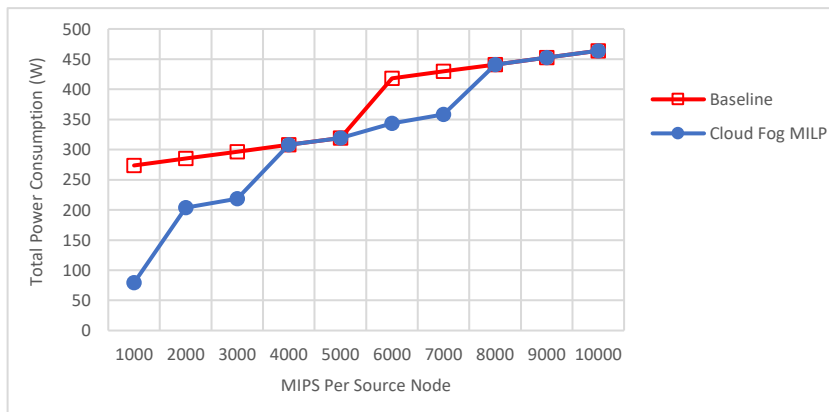

**Figure 12** Total power consumption in Scenario three of the cloud fog approach vs. the baseline approach.

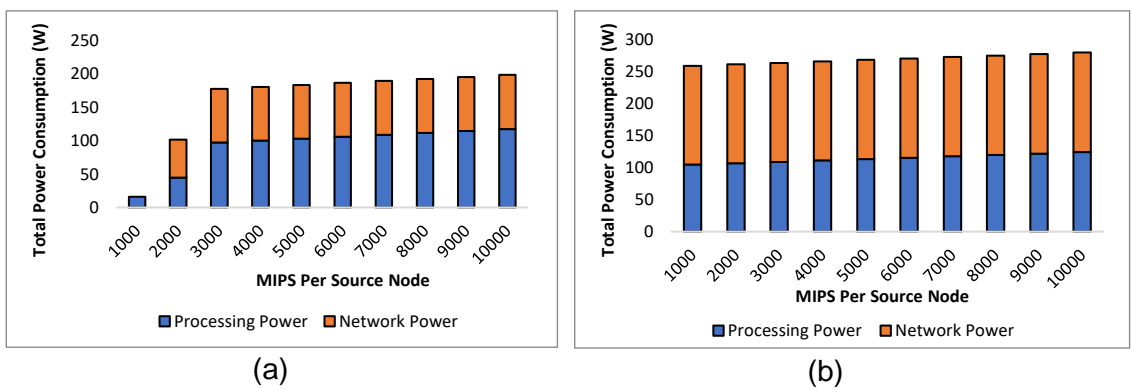

**Figure 13** Total power consumption in Scenario four broken down into network and processing power consumption in a) cloud fog approach and b) baseline approach.

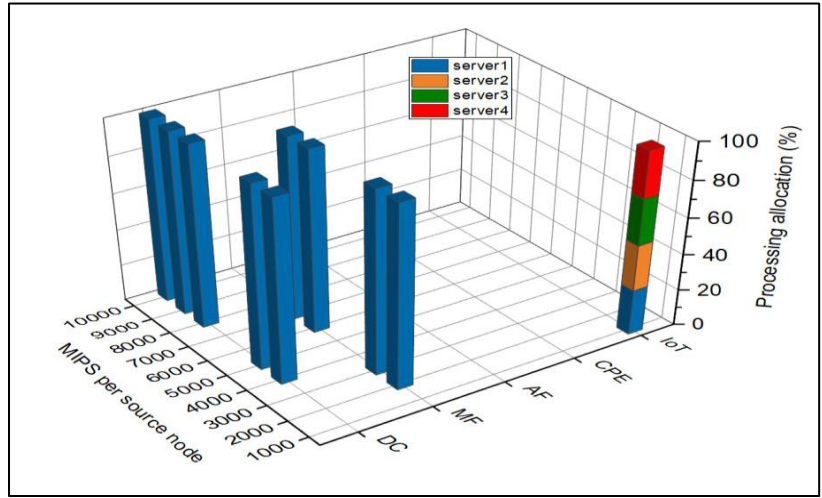

**Figure 14** Processing task allocation.

Energy & Delay Aware Processing Placement

In this section, we study the trade-off between power consumption and delay. We consider propagation delay between network nodes and queuing delay at different network nodes. We optimize the allocation of the processing resources in a multi-objective MILP optimization model to minimize the power consumption and delay equally.

The propagation delay is based on the distances between network nodes as this network covers a large geographical area, and is calculated using the following equation:

$$Propagation\ Delay = \frac{D}{C} \quad (41)$$

where $D$ represents the distance, and $C$ is the speed of light.

The values of distance (D) between each two nodes are based on the following assumptions:

1. The distance between the AP and surrounding IoT nodes is set to the standard coverage range of WLAN, (100 m) [71].
2. The distance between the AP and ONU is estimated based on an assumption that one ONU can connect to multiple APs in a typical LAN (100 m) [72].
3. The distance between the ONU and OLT is based on typical PON designs. We considered a design where the OLT is located in the telecom main office in the centre of the city. ONUs usually represent devices located at the end-users location (i.e., at home); usually such distance are around 5–20 km [73] so we assumed an average distance equal to (10 km).
4. The distance between the OLT and metro node (router and switch) was estimated based on the metro network design. The metro network usually has a radius of 20–120 km [74]. The OLT can be either collocated with the metro node in the same telecom office or located somewhere else in the local area of the metro node (1–10 km away). We based our estimation on the latter scenario with an approximate distance equal to (5 km) between the OLT and metro node.
5. The distance between the metro node and the core node (including the associated data centre), is given as the distance between two large cities, assuming the current city does not have a large central cloud. An example of such a distance is taken as the distance between Leeds and a large data centre in London, a (300 km) distance.

The queuing delay was modelled for each networking node as an M/M/1 queue with one server, where arrivals follow a Poisson process and the service rate is negative exponentially distributed, summarized in Figure 15. The queueing delay was calculated based on aggregated traffic at each node (arrival rate) and the maximum capacity of the nodes (service rate), as given below

$$Queuing\ Delay = \frac{1}{\mu - \lambda} \quad (42)$$

where $\mu$ is the service rate, and $\lambda$ is the arrival rate. We have considered in this work delay at the packet level. We used the Ethernet maximum packet size of 1500 bytes and therefore, expressed the arrival data rates as packets per second and expressed the service rates (transmission rates) in packets per second.

Three different service rate values are considered. We assumed that the AP works based on the wireless medium interface capacity (with 1 Gb/s service rate). Moreover, the core node, with the associated data centre, was assumed to work at 40 Gb/s, as they are part of the IP/WDM network. Other network devices were assumed to have a 10 Gb/s service rate based on GPON.

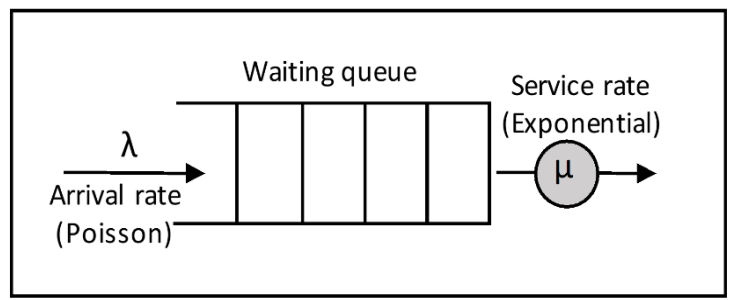

**Figure 15** M/M/1 Queueing model

The MILP model introduced in the previous sub-section was extended to jointly minimize power consumption and delay. To continue to use linear programming, Equation *(42)* was converted to a linear form based on a pre-defined lookup table. This table includes all the possible generated traffic combinations (arrival rates indicator), indexed with the calculated queuing delay based on a fixed service rate. As we have three different service rate values in our designed network, three lookup tables were defined. Based on this arrival rate indicator, the queuing delay for a node was given as the value corresponding to the indicator in the lookup table.

The modified MILP defines the following additional sets, parameters, and variables:
**Sets:**
$\mathbb{AR}$     Set of arrival rates.

$\mathbb{SR}$     Set of service rates.

**Parameters:**
$\eta_{as}$     Queuing delay at arrival rate $a \in AR$ and service rate $\in SR$, in the lookup table.

$G1$     Large enough number with units of Mb/s.

$G2$     Large enough number with units of ms.

$D_{mn}$     Distance between any two nodes $(m, n)$, where $i \in \mathbb{N}, j \in \mathbb{N}_m$.

$C$     Speed of light, $C = 299{,}792 \frac{km}{s}$.

| | |
|---|---|
| $\Delta RI$ | Refractive index of fibre, which is defined as the ratio of the speed of light in fibre to speed of light in free space; $\Delta RI = \frac{2}{3}$. |

**Variable**

| | |
|---|---|
| $\zeta_{mn}^{sd}$ | Binary variable $\zeta_{ij}^{sd} = 1$ if the traffic flow sent from source node $s$ to processing node $d$ traverses physical link $(i,j)$, where $s \in \mathbb{S}$, $d \in \mathbb{P}$, and $m, n \in \mathbb{N}$. |
| $Q_{mn}^{sd}$ | Queuing delay at node $j$ experienced by the traffic from source node $s$ to processing node $d$ traversing physical link $(i,j)$, where $s \in \mathbb{S}$, $d \in \mathbb{P}$ and $i, j \in \mathbb{N}$. |
| $Q_i$ | Queuing delay experienced by traffic aggregated at node $i \in \mathbb{N}$. |
| $Q_{sd}$ | Queuing delay of the traffic sent from source node $s \in \mathbb{S}$ to processing node $d \in \mathbb{P}$. |
| $Q$ | Total queuing delay of the network. |
| $R_{sd}$ | Propagation delay of the traffic sent from source node $s \in \mathbb{S}$ to processing node $d \in \mathbb{P}$. |
| $R$ | Total propagation delay of the network. |
| $\lambda_i$ | Arrival rate (total traffic) at each node $i \in \mathbb{N}$. |
| $H_{ij}$ | Arrival rate indicator for node $i \in \mathbb{N}$, $\sigma_{ij} = 1$ if the arrival rate of node $i$ matches rate $j \in AR$, it is 0 otherwise. |

All the power consumption equations in the previous sub-section were considered in this model. The total power consumed in is calculated as follows:

$$\text{Total Power Consumption} = net\_pc + pr\_pc \quad (43)$$

Additionally, the following equations are used to calculate the propagation and queuing delay for the network.

1) The total propagation delay (R), is calculated based on the propagation delay between all source node and processing node pairs and is given as

$$R = \sum_{s \in \mathbb{S}} \sum_{d \in \mathbb{P}} R_{sd} \qquad \forall \ s \in \mathbb{S}, d \in \mathbb{P} \quad (44)$$

where $R_{sd}$ is is the propagation delay of the path traversed by traffic sent from each source node $s \in \mathbb{S}$ to the processing node $d \in \mathbb{P}$, and is calculated as follows;

$$R_{sd} = \sum_{\substack{m \in \mathbb{N} \\ m \notin \mathbb{I}}} \sum_{n \in \mathbb{N}_m} \zeta_{mn}^{sd} \frac{D_{mn}}{\Delta RIC} \qquad \forall \ s \in \mathbb{S}, d \in \mathbb{P} \quad (45)$$

$$R_{sd} = \sum_{\substack{i \in \mathbb{N} \\ i \in \mathbb{I}}} \sum_{j \in \mathbb{N}_m} \zeta_{mn}^{sd} \frac{D_{mn}}{\mathbb{C}} \qquad \forall \ s \in \mathbb{S}, d \in \mathbb{P} \quad (46)$$

Equation *(45)* and *(46)* calculate the propagation delay for the traffic sent to the processing nodes via fibre or wireless links, respectively. A refractive index $\Delta RI$ with the value of $\frac{2}{3}$ is added to Equation *(45)* to define the ratio of the speed of light in fibre to the speed of light in free space.

2) The total queuing delay (Q), which is calculated based on the queuing delay experienced by traffic between all the source node and processing node pairs, and is given as:

$$Q = \sum_{s \in \mathbb{S}} \sum_{d \in \mathbb{P}} Q_{sd} \quad \forall \ s \in \mathbb{S}, d \in \mathbb{P} \tag{47}$$

where $Q_{sd}$ is the queuing delay of the path traversed by traffic sent from each source node $s \in \mathbb{S}$ to processing node $d \in \mathbb{P}$, and is calculated as:

$$Q_{sd} = \sum_{m \in \mathbb{N}} \sum_{j \in \mathbb{N}_m} Q_{mn}^{sd} \quad \forall \ s \in \mathbb{S}, d \in \mathbb{P} \tag{48}$$

Equation *(48)* calculates the queuing delay for a traffic demand by summing the queuing delay experinced by the demand at each node.

The joint objective is defined as: N

**Minimize**

$$\alpha P + \beta R + \gamma Q \tag{49}$$

where $\alpha$, $\beta$, and $\gamma$ are weight factors used for the following purposes: (i) to scale the terms so that they are comparable in magnitude; (ii) to emphasise and de-emphasise terms (power, queuing delay and propagation delay); and (iii) to accommodate the units in the objective function. Therefore, $\alpha$ is a unitless factor, and $\beta$ & $\gamma$ have units of $\frac{Watt}{sec}$.

In addition to the constraints in the previous sub-section, the model is subject to the following additional constraints:

1) The traffic estimation at each node:

$$\sum_{s \in \mathbb{S}} \sum_{d \in \mathbb{P}} \sum_{m \in \mathbb{N}_i} \lambda_{nm}^{sd} = \lambda_m \quad \forall \ m \in \mathbb{N}, m \notin \mathbb{S} \tag{50}$$

Constraint *(50)* calculates the traffic arrival at each node in the.

2) The arrival rate indicator:

$$\sum_{n \in \mathbb{AR}} H_n = \lambda_m \quad \forall \ m \in \mathbb{N}, m \notin \mathbb{S} \tag{51}$$

Constraint *(51)* creates indicators of the arrival rate for each node. This is equal to 1 if the arrival rate is equal to $n$:

$$\sum_{n \in \mathbb{AR}} H_{mn} \leq 1 \quad \forall \ m \in \mathbb{N}, m \notin \mathbb{S} \tag{52}$$

Constraint *(52)* ensures that each node has no more than one arrival rate indicator for a given service rate.

3) Queuing delay estimation:

$$\sum_{n \in \mathbb{AR}} H_{mn} \cdot \eta_{ns} = Q_m \qquad \forall\ m \in \mathbb{C} \cup \mathbb{DC},\ s = 40Gb/s \tag{53}$$

$$\sum_{n \in \mathbb{AR}} H_{mn} \cdot \eta_{ns} = Q_i \tag{54}$$
$$\forall\ m \in \mathbb{N}, m \notin \mathbb{I} \cup \mathbb{C} \cup \mathbb{DC} \cup \mathbb{AP},\ s = 10Gb/s$$

$$\sum_{n \in \mathbb{AR}} H_{mn} \cdot \eta_{ns} = Q_i \qquad \forall\ m \in \mathbb{AP},\ s = 1Gb/s \tag{55}$$

Constraints *(53)* to *(55)* estimate the traffic delay for each node that operates at 40 Gb/s, 10 Gb/s, or 1Gb/s respectively.

$$\lambda_n^{sd} \geq \zeta_{mn}^{sd} \qquad \forall\ s \in \mathbb{S}, d \in \mathbb{P}, m\ and\ n \in \mathbb{N} \tag{56}$$

$$\lambda_{ij}^{sd} \leq G1\ \zeta_{ij}^{sd} \qquad \forall\ s \in \mathbb{S}, n \in \mathbb{P}, i\ and\ j \in \mathbb{N} \tag{57}$$

Constraints (56) and (57) set $\zeta_{mn}^{sd} = 1$ if the traffic demand between the source node and the processing node is routed throughlink $(m, n)$.

$$Q_{mn}^{sd} = Q_n\ \zeta_{mn}^{sd} \qquad \forall\ s \in \mathbb{S}, n \in \mathbb{P}, m\ and\ n \in \mathbb{N} \tag{58}$$

$$Q_{mn}^{sd} \leq G2\ \zeta_{mn}^{sd} \tag{59}$$

$$Q_{mn}^{sd} \leq Q_n \tag{60}$$

$$Q_{]mn}^{sd} \geq Q_n - G2\left(1 - \zeta_{mn}^{sd}\right) \tag{61}$$

Equation *(58)* calculates the queuing delay at node $m$ for the traffic sent from source node $s$ to processing node $d$. As Equation *(58)* involves the multiplication of two variables, $Q_{ij}^{sd}$ and $Q_i$, constraints *(59)* to *(61)* are used to remove the nonlinearity of Equation *(58)* and replace the relationship with an equivalent linear relationship.

## Scenarios and Results

The model presented in the previous section is considered with the following variations of the objective function:

1) Minimizing the total power consumption only

2) Minimizing the traffic propagation delay only

3) Minimizing the power consumption and traffic propagation delay jointly.

4) Minimizing the traffic queuing delay

5) Minimizing the power consumption and traffic queuing delay jointly.

6) Minimizing the power consumption, traffic propagation and queuing delay jointly.

The previously described objective functions were combined into four evaluations that highlight the individual effects of the propagation and queuing delay, combined with the power consumption on the processing allocation decision, and both power and delay values. All the evaluations consider Scenario Two described in the previous sub-section, where we considered a cloud-fog-VEC allocation (CFVA) with low-density VNs (8VNs) and single five tasks were generated from the same IoT group. Each source node in group 1 generates a task with an increasing preocessing requirement (1000–10000 MIPS) and a propotional data rate ranging from 1Mbps to 10 Mbps per task.

## 1) Evaluation One: Power and Propagation Delay Minimization

In this evaluation, we study the joint minimization of the power consumption and the propagation delay (objective function case 3), and compare the results to the two cases where only the power (objective function case 1) or propagation delay (objective function case 2) are minimized. Figure 16 and Figure 17 illustrate the total power consumption and the average propagation delay for the three cases considered in the objective function versus the traffic generated per demand, (1–10) Mbps. These results are reflected by the processing allocation illustrated in Figure 18.

Figure 16 shows that the three objective functions have achieved the minimum power consumption when processing tasks locally in the IoT nodes. Local processing is confirmed to be the most energy efficient strategy (per the results generated from the previous sub-section). Moreover, it is confirmed that local processing can achieve minimize propagation delay as multiple hops and distances are avoided, as shown in Figure 18. The propagation delay minimized case, in Figure 17 produced the highest power consumption. The jumps in the curve (at 2 Mbps) are due to moving the allocation to a less efficient PN that can support the demand. Moreover, activating two processing nodes, as shown in Figure 18, causes an increase in the power consumption compared to the power minimization case. This increase continues as all tasks are allocated to the access fog by activating all its servers compared to one metro fog server in the power minimization case. With the joint minimization of the power and propagation delay, results led to a lower power consumption, by an average of 40%, compared to the delay minimization objective, as shown in Figure 17. However, the propagation delay, in Figure 18, increases at high traffic because the model activated a metro server and access server instead of two access servers in the delay minimization case. This is to achieve a balance between the power consumption and propagation delay. Activating an access server decreases the propagation delay but metro servers minimize the processing power consumption and therefore the total power consumption, as seen in Figure 17.

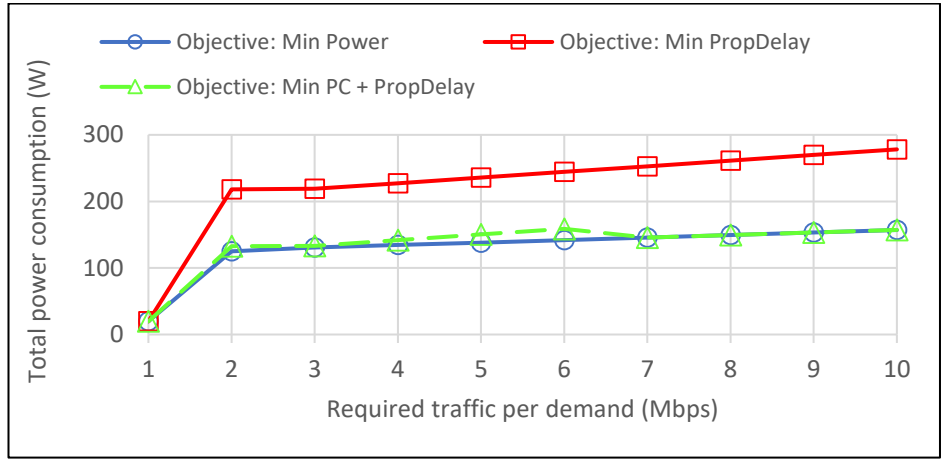

Figure 16 Total power consumption (power and propagation delay minimization)

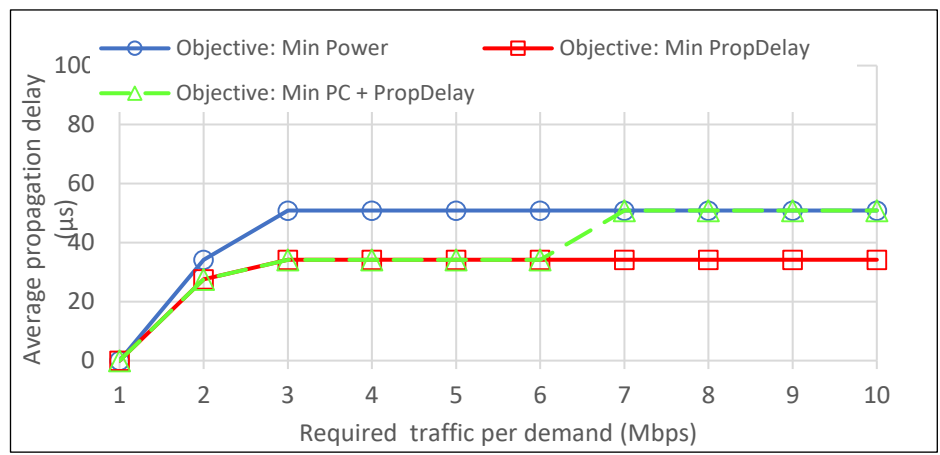

Figure 17 Average propagation delay (power and propagation delay minimization)

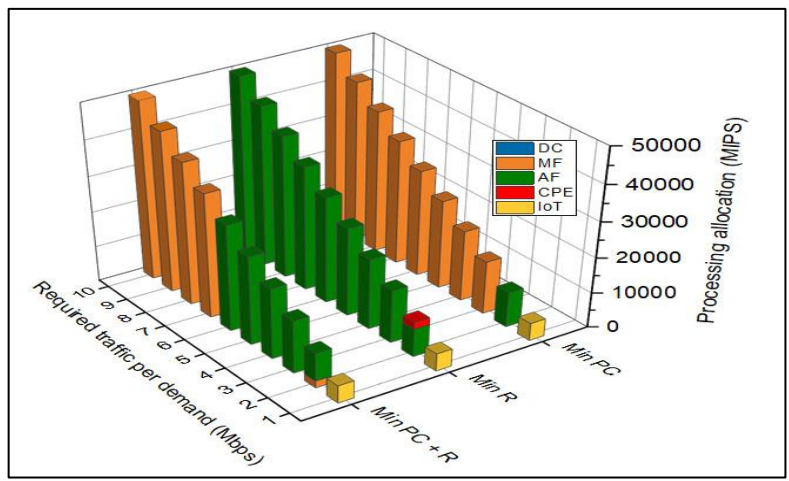

Figure 18 Processing allocation at each processing node (power and propagation delay minimization).

## 2) Evaluation Two: Power and Queuing Delay Minimization

In this evaluation, we study the joint minimization of the power consumption and the queuing delay (objective function case 5), and compare it to the power minimized case (objective function case 1), and the queuing delay minimized objective function (objective function case 4). Figure 19 shows the total power consumption for the three cases. Similar to the propagation delay minimization case, minimizing queuing delay consumes the highest power consumption. This is due to allocating tasks to the processing nodes that guarantee the minimum hops and therefore minimum queuing delay experienced by each networking node. For example, as seen in *Figure 21*, tasks are allocated to the CPE whenever it is sufficient. Subsequently, all tasks are allocated to the AF by activating all the servers in the access fog. On the other hand, relatively comparable power consumption results can be observed for the power minimized case (blue curve) and power and delay minimized case (yellow curve), in Figure 19. The latter case causes more power consumption. With 3-4 Mbps generated traffic allocating low traffic tasks to AF achieves a balance between the power consumption and queuing delay. Figure 19 shows comparable average queuing delay for the three minimization cases. This is due to the fact that all the networking nodes in the access and metro layers operate with the same service rate. The increase in the queuing delay in the power minimized cases is caused by the extra hop the traffic travels through when allocating the tasks to the metro layer, as shown in *Figure 21*.

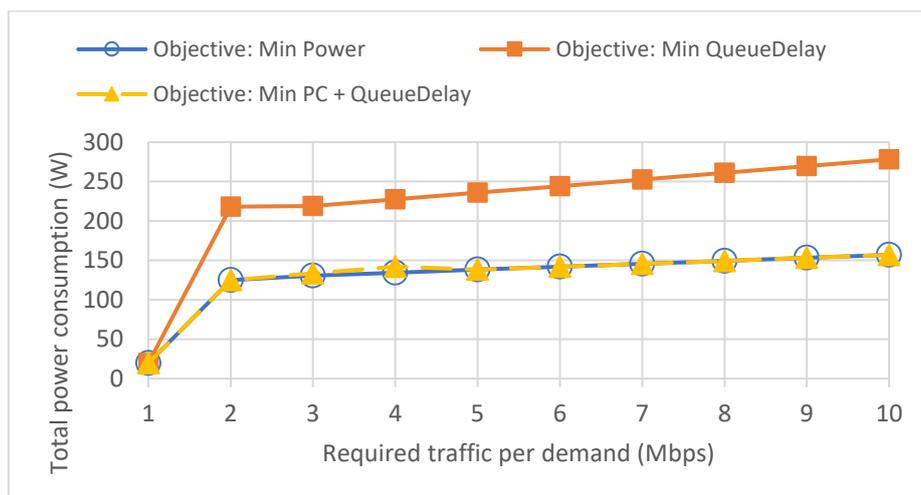

*Figure 19 Total power consumption (power and queuing delay minimization)*

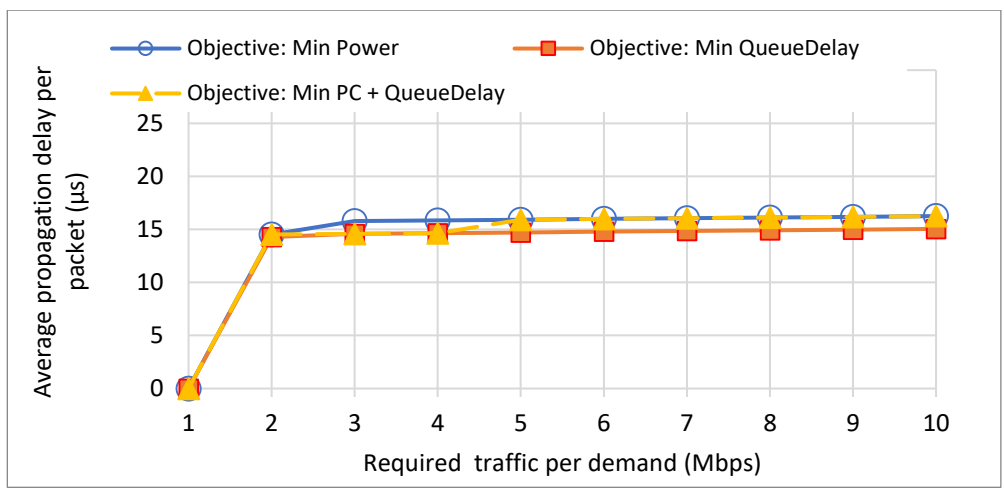

Figure 20 Average queuing delay per packet (power and queuing delay minimization)

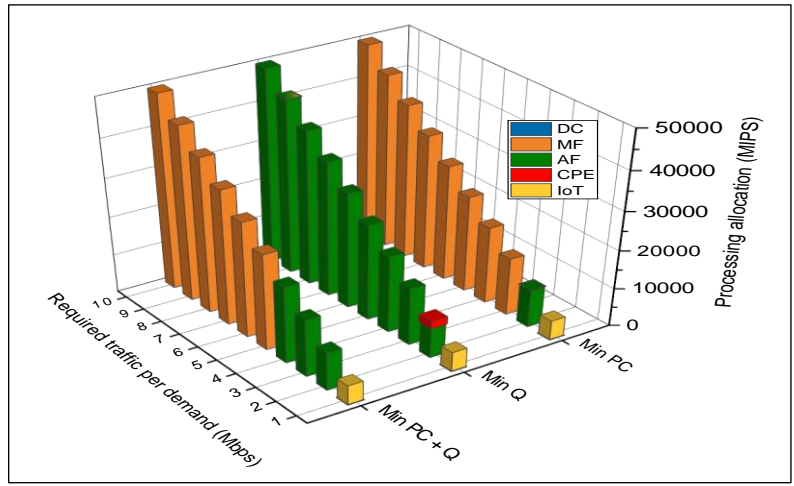

Figure 21 Processing allocation at each processing node (power and queuing delay minimization)

# Conclusions and Future Work

In this chapter, we have evaluated a cloud fog architecture for future 6G networks paying special attention to energy efficiency and latency. We developed a Mixed Integer Linear Programming (MILP) model that is generic and independent of technology and application. We used the resultant model to investigate the processing task allocation problem in a representative IoT application. The results showed that in the cloud fog approach despite its limitations such as processing resources and distributed nature, substantial amounts of power savings can be achieved. The results also showed that regardless of how efficient edge processing is, the cloud DC will always remain relevant due to its abundance of processing capacity and efficiency. We have also investigated the joint optimization of power consumption, propagation delay and queueing delay when allocating processing tasks to the available servers. Three evaluations were considered with different objective functions where power consumption is examined with propagation delay and queueing delay. Our results show that the closer the server is to the source nodes, the lower the propagation and queuing delay achieved, as the distance and the number of hops affect the propagation delay and queuing delay. However, the queuing delay can be reduced by utilizing higher data rates in the networking devices. Therefore the effects in terms of increased delay when allocating tasks to a further location can be reduced by using higher data rates on route, and depend on the number of hops the traffic traverses.

Future work can introduce additional optimization components to the delay such as processing and transmission delay alongside propagation and queuing delay. It can also consider queuing at the wireless devices where IoT nodes are allowed to communicate to each other. Future studies can also include developing heuristic algorithms to mimic the behaviour of the MILP models as well as accounting for the dynamic nature of demands in uncertain network settings. It is also worth looking into duty cycling schemes (shallow and/or deep sleep) which can result in further power savings. Also, it is it has to be observed that shutting down a network or processing element completely poses latency challenges in the start-up phase.